\def\year{2021}\relax
\documentclass{article}
\usepackage[utf8]{inputenc}
\usepackage{aaai21}
\usepackage{times}
\usepackage{helvet}
\usepackage{courier}
\usepackage[hyphens]{url}
\usepackage{graphicx}
\usepackage{natbib}
\usepackage{caption}
\usepackage[noend]{algpseudocode}
\usepackage{algorithm}
\usepackage{algorithmicx}
\usepackage{bbding}
\newcommand{\system}{Oscars}

\title{\system{}: Adaptive Semi-Synchronous Parallel Model for Distributed Deep Learning with Global View}

\author{Sheng Huang}
\par
\affiliations{
    \textsuperscript{\rm 1} Shanghai University of Electronic Power
}

\begin{document}
\maketitle

\begin{abstract}
Deep learning has become an indispensable part of life, such as face recognition, NLP, etc., but the training of deep model has always been a challenge, and in recent years, the complexity of training data and models has shown explosive growth, so the training method is gradually transformed into distributed training. Classical synchronization strategy can guarantee accuracy but frequent communication can lead to a slow training speed, although asynchronous strategy training speed but can not guarantee the accuracy, and in the face of the training of the heterogeneous cluster, the above work is not efficient work, on the one hand, can cause serious waste of resources on the other hand, frequent communication also made slow training speed, so this paper proposes a semi-synchronous training strategy based on local-SDG, effectively improve the utilization efficiency of heterogeneous resources cluster and reduce communication overhead, to accelerate the training and ensure the accuracy of the model.
\end{abstract}
\section{Introduction}
\label{intr}

It has been widely acknowledged that machine learning has become fundamentally important in a wide range of research and engineering areas, including autonomous driving, face recognition, speech recognition \cite{deng2013recent}, text understanding \cite{lang/mikolov2013efficient,lang/liang2017recurrent}, image classification \cite{image/yan2019hierarchical,image/yan2016automatic}, etc. There has been an imperative
need to improve the performance when training machine
learning models, especially in the presence of larger volumes
of data and increasingly complex computing models. Current the structure of neural network have  at hundreds layers are relatively common, such as the Bert\cite{devlin2018bert} language model proposed by Google contains 300 million parameters, ImageNet\cite{deng2009imagenet} data set contains 20000 categories a total of 15 million images. 

Parameter servers\cite{li2013parameter}.  are widely used in today's distributed training system, such as MXNet\cite{chen2015mxnet} and TensorFlow\cite{abadi2016tensorflow}.The architecture is shown in the Fig.1.
Parameter server architecture consists of a logic server group and a lot of workers. Under the parameter server architecture. Each worker holds different training data and the same copy of the model. Each worker calculates the gradient locally, then periodically push the local gradient to parameter server, and then parameter server summarizes the gradient of each worker and updates the model parameters. Finally, each worker pull the latest parameters to continue the training. In addition, there is a decentralized architecture that is distinct from 
Parameter server, called Ring-AllReduce. In this architecture, all nodes form a logic ring, and each node only communicates with its neighbor nodes, effectively avoiding the bandwidth congestion caused by centralization. However, due to the characteristics of the architecture, only synchronous algorithm can be used, so Straggler in this architecture will lead to more serious problems. At present, there are also some work to optimize ring, such as horovod
\cite{horovod,gibiansky2017bringing,jia2018highly,mikami2018imagenet}. In this work, we focus on parameter server.

At present, the prevalent synchronous paradigms are: Bulk Synchronous Parallel(BSP)\cite{gerbessiotis1994direct}, Asynchronous Parallel (ASP), and Stale Synchronous Parallel (SSP)\cite{ho2013more}. 
BSP, as a famous general parallel computing synchronization model in distributed computing. 
 Due to its stability and reliability, it has the same stability as SGD on a single machine. Thus the mainstream distributed training systems take it as the default parallel strategy. Barriers are a critical component in BSP. It requires each node to stop working after completing its own tasks and to wait until all nodes have completed their tasks. Although this can ensures a high degree of consistency of models on different nodes, it has serious drawbacks. In a heterogeneous or volatile cloud environment, the performance of each node is not the same. This means that each node takes different amounts of time to process the same amount of data, which results in a large amount of time spent waiting for the slowest node in each synchronization process. A typical example is shown in Figure 2. 
 \begin{figure}[tp]
  \centering
  \includegraphics[width=0.45\textwidth]{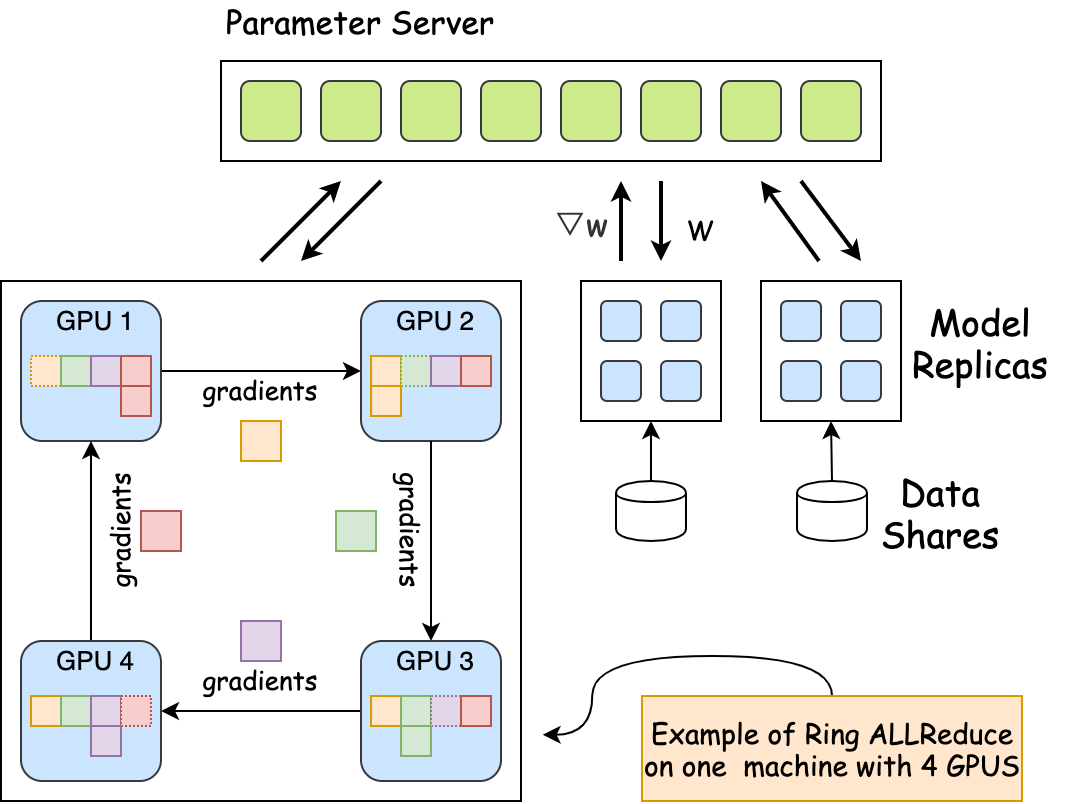}
  \caption{The architecture of Parameter Server and Ring-AllReduce.}
  \label{ps}
\end{figure}

 So a natural idea would be to relax synchronization requirements.
 \begin{figure}[tp]
  \centering
  \includegraphics[width=0.5\textwidth]{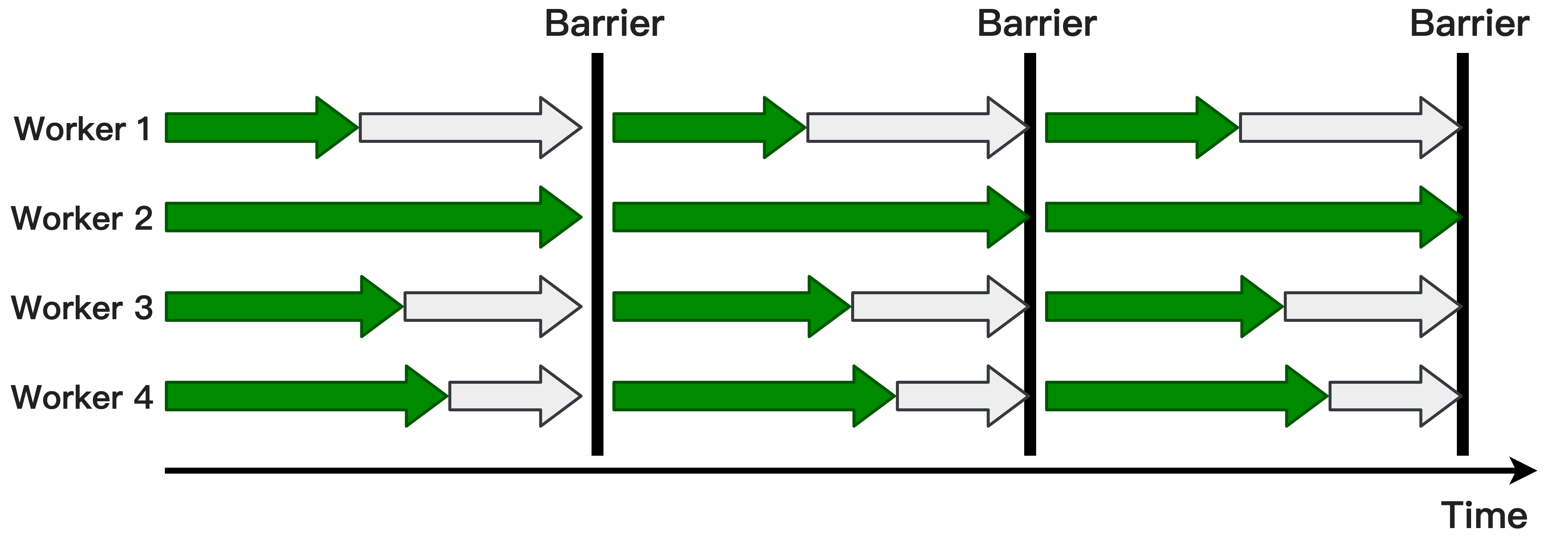}
  \caption{Under the restriction of barrier, the fastest node must wait for the slowest node after each iteration, which is very inefficient}
  \label{bspli}
\end{figure}
 According to this naive idea, ASP paradigms remove the strict barriers, so that each node can work asynchronously.
However, because the progress of each node is too different, the model will shake, and it will take longer time to converge. 
SSP\cite{ho2013more} is a compromise between the two methods. As long as the pace between the fastest node and the slowest node does not exceed the stale threshold, each node can work asynchronously. But due to SSP introduces an extra threshold,If the threshold setting is not reasonable, it will also be seriously affected by the straggler, causing the fastest nodes frequently to stop waiting for the slowest node during the training process or too big causing non-converge.

In order to make up for the shortcomings of the above  Parallel paradigm. We propose a adaptive load balance approach. The aim of our approach is to relax the strict synchronization requirement of classical BSP and to improve resource utilization. The core idea is let slower workers do less computation between synchronization, and faster workers do more. Under global control, the waiting time of each synchronization is minimized. Thus, the overall training time is shortened.  Our contributions are summarized as follows:

\section{Motivation}
\label{mo}



\begin{figure}[ht]
    \centering
    \includegraphics[width=1\columnwidth]{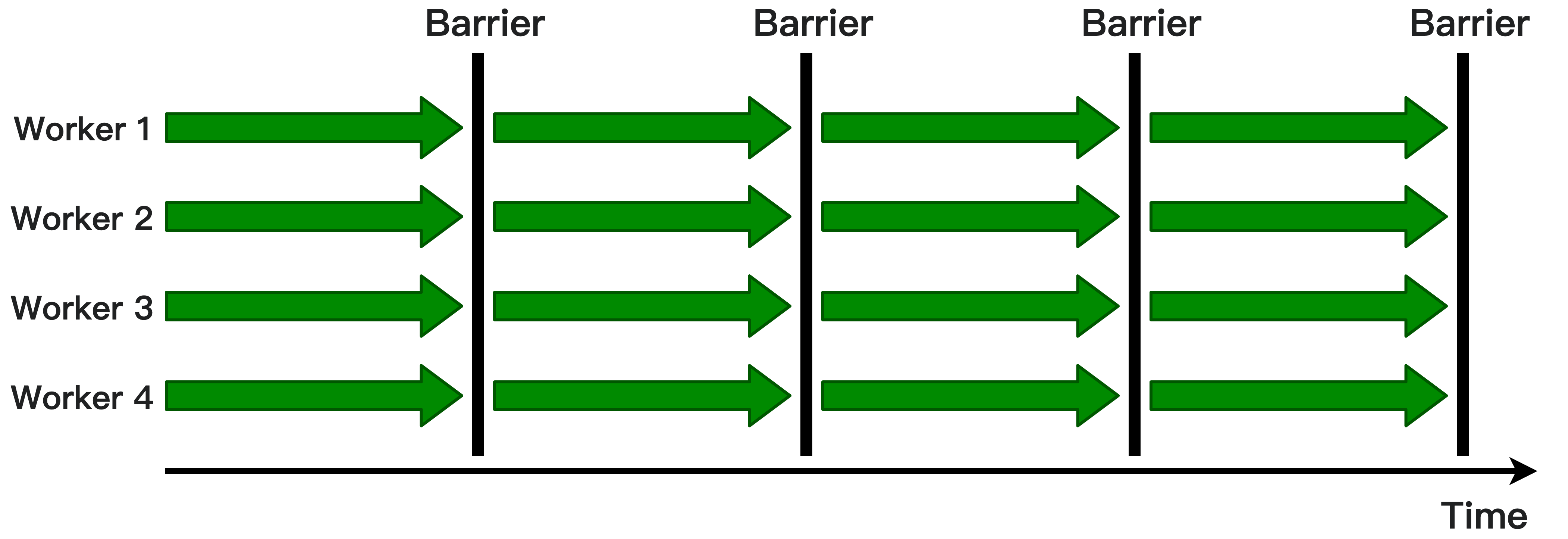}
    \caption{Mini-batch SGD on homogeneous environment. The green arrow represent the computation.}
    \label{bsp_on_hom}
\end{figure}

Mini-batch stochastic gradient descent (SGD) is state of the art in large scale distributed training (See Figure \ref{bsp_on_hom}). The scheme can reach a linear speedup with respect to the number of workers, but this is rarely seen in practice as the scheme often suffers from large network delays and bandwidth limits. To overcome this communication bottleneck recent works propose to reduce the communication frequency. An algorithm of this type is Local SGD \cite{zinkevich2010parallelized,mcdonald2010distributed,zhang2014improving,mcmahan2017communication} that runs SGD independently in parallel on different workers and averages the sequences only once in a while.

\begin{figure}[ht]
    \centering
    \includegraphics[width=1\columnwidth]{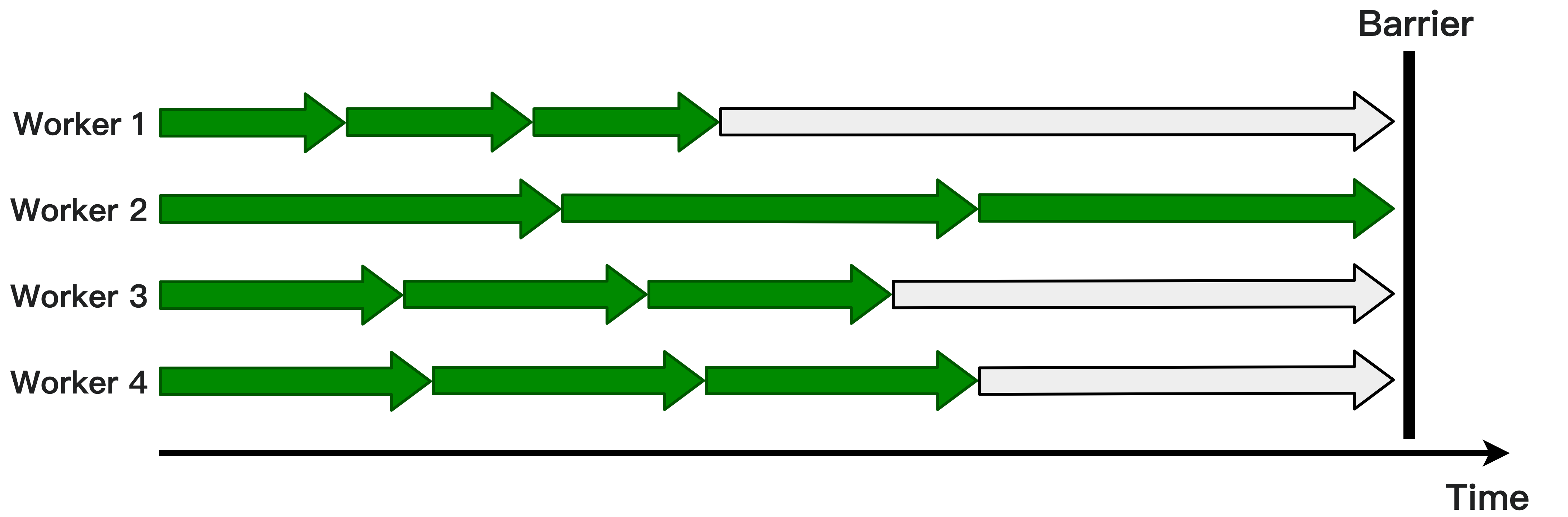}
    \caption{Local SGD in heterogeneous environment. The green arrow represent the computation and the gray arrow represent the idle state.}
    \label{local_without_balancing}
\end{figure}

Local SGD requires all workers to compute the average of individual solutions every I iterations and synchronization among local workers are not needed before averaging. However, the fastest worker still needs to wait until all the other workers finish I iterations of SGD even if it finishes its own I iteration SGD much earlier. (See Figure \ref{local_without_balancing} for a 4 worker example where one worker is significantly faster than the others.) As a consequence, the computation capability of faster workers is wasted. Such an issue can arise quite often in heterogeneous networks where nodes are equipped with different hardwares.

In this paper, we present asynchronous local SGD with load-balancing (Figure \ref{local_with_balancing}) that does not require that the local sequences are synchronized. This does not only reduce communication bottlenecks, but by using load-balancing techniques the algorithm can optimally be tuned to heterogeneous settings (slower workers do less computation between synchronization, and faster workers do more).


\begin{figure}[ht]
    \centering
    \includegraphics[width=1\columnwidth]{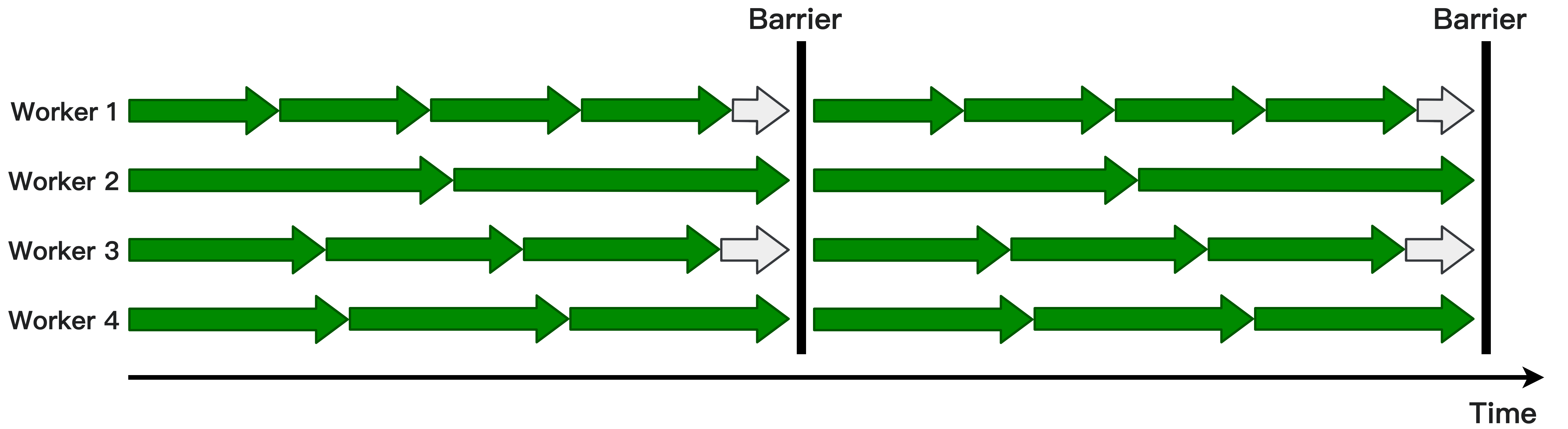}
    \caption{Local SGD with load-balancing in heterogeneous environment. The green arrow represent the computation and the gray arrow represent the idle state.}
    \label{local_with_balancing}
\end{figure}

\section{Problem Formulation  }
\label{pf}

Most data centers have high availability, assuming that they are run in a stable environment, and each worker's computational speed is in a steady state.

According to the case study, the time spent by a worker before the global barrier can be composed of three parts, the barrier notated as $T$. The first part is the gradient calculation time, notated as $t_{i}^{iter}$. The second part is the idle time of wait for other workers to synchronize parameters, notated as $t_{i}^{w}$.  and the third part is the time of synchronize the parameters. We assume that the bandwidth between workers in the data center is very fast, so we will ignore the third part.

For N workers in a heterogeneous cluster, each worker has to process a certain amount of iterations before the global barrier where each iteration time on the same worker is similar.
Assume N worker index by $i$, the time of a local iteration of each worker can be notated as $t_{1}^{iter}, t_{2}^{iter}, t_{3}^{iter}...t_{N}^{iter}$. Given a global barrier $T$, we can get the $t_{i}^{w}$ of each worker:
\begin{equation}
   t_{i}^{w} = mod(T,t_{i}^{iter})
\end{equation} 

Before a synchronization, the maximum waiting time can reflect the idle degree of workers. If the maximum wait time is as small as possible, it means that all worker can complete the last batch calculation  exactly at the global barrier. Thus, computing resources can be fully utilized. We define the maximum wait time as:
\begin{equation}
      max(mod(T, t_{i}^{iter}))
      \label{max_wait_time}
\end{equation} 
 
 So, We are looking for the optimal $T^{*}$ which gives the minimum Eq. \ref{max_wait_time} from all possible $T$. At last, we formulate the following optimization problem: 
\begin{equation}
    T^{*} = \mathop{argmin}\limits_{T} max mod(T, t_i)
\end{equation} 
$  S.T. \quad  floor(T/min(t_{i}^{iter})) - floor(T/max(t_{i}^{iter})) < M $
Where $floor(T/(t_{i}^{iter}))$ represent how many times can the $i$-th worker iterate before the barrier at most, notated as $\tau_{i}$, and $M$ limits the difference of local step of different workers in the appropriate range.


\begin{table}[]
  \begin{tabular}{|l|l|}
  \hline
  
  Name & Description \\ \hline
  $t_{i}^{iter}$  & Time of a local iteration of worker i \\ \hline
    $t_{i}^{w}$  & Time of wait of worker i \\ \hline
  $T$  &  The time point of global synchronization \\ \hline
  $N$  &  Number of workers \\ \hline
  \end{tabular}
  \caption{Frequently used notations}
\end{table}

\section{Approach}
\label{ap}

In this section we present load-balance local SGD. This does not only using load-balancing techniques which the algorithm can optimally be tuned to heterogeneous settings (slower workers do less computation between synchronization, and faster workers do more), but also reduce the network overhead caused by the frequently communication.

\subsection{Local-SGD load balancing}
To minimize the wait time and improve cluster utilization, we propose an fast and efficient algorithm based on the principle of least common multiple. The algorithmic flow can be found here\ref{alg:Load-Balance Algorithm},and the described as follows: First let  $T=max(t_{i}^{iter})$. Next use a loop to get the max value of $mod(T,t_{i}^{iter})$, and $T++$. The above loop is repeated until the constraint is not satisfied. Last take the $T^{*}$ that make the max value of $mod(T,t_{i}^{iter})$ is the minimum. Therefore, the total computation complexity is $O(MN)$ which is an linear complexity. It will not bring additional overhead to the original training system.

After obtain the optimal $T^{*}$, we can calculate the number of iterations for each worker according to $t_{i}^{iter}$. The number of iterations of each worker is expressed as: $\tau_{1},\tau_{2}...\tau_{N}$. So the model update rule is:

\begin{equation}
 x_{t+1}^{i} = \left\{ \begin{array}{ll}
\frac{1}{N} \sum_{k=1}^{N} (x_{t}^{k} - \eta g(x_{t}^{k}))  & \textrm{ $t \ mod \ \tau_{i} =0$}\\
x_{t}^{i} - \eta g(x_{t}^{i})& \textrm{otherwise}
\end{array} \right.
\end{equation}

where $x_{t}^{i}$ denotes the model parameters in the $i$-th worker.

\subsection{Data partition load balancing}

\begin{algorithm}
\caption{Load-Balance Algorithm}
\label{alg:Load-Balance Algorithm}
\begin{algorithmic}

\State {\textbf{Input: } $t_{i}^{iter},M$}
\State {\textbf{Output: } $T^{*}$}
\State {\textbf{Initial: }$T=max(t_{i}^{iter})$}

\For{$floor(T/min(t^{iter})) - floor(T/max(t^{iter})) < M , i=1$}
    \For{$each\quad worker:j=1,2,3...N$}
    \State {$t^{w}[j] = mod(T,t_{j}^{iter})$}
    \EndFor
    \State $max\_wait\_time[i]$ = $max(t^{w})$
    \State $T\_set[i] = T$
    \State $T++$
\EndFor
$T^{*} = T\_set[indexOf(min(max\_wait\_time))]$

\end{algorithmic}
\end{algorithm}

\section{Evaluation Setup}
\subsection{Testbed}
We conduct our experiments on a GPU server. The server runs with 2 NVIDIA RTX 2080 GPUs and interconnected with 10Gbps PCI-E. The server run Ubuntu Server 18.06. We used Pytorch framework to build our algorithm prototypes.

\subsection{Dataset and DL Models}
We used CIFAR-10 datasets for image classification tasks. The datatsets has 50,000 training images and 10,000 test images. We choosed ResNet101 as our deep neural network baseline to evaluate our approach.

\subsection{Metrics}
The performance metrics include scalability and Rate of convergence. The scalability denotes the speedup on throughput (number of iterations finished per hour) compared with single node DL.

\section{Evaluations}
\label{exp}

\begin{figure}[ht]
    \centering
    \includegraphics[width=1\columnwidth]{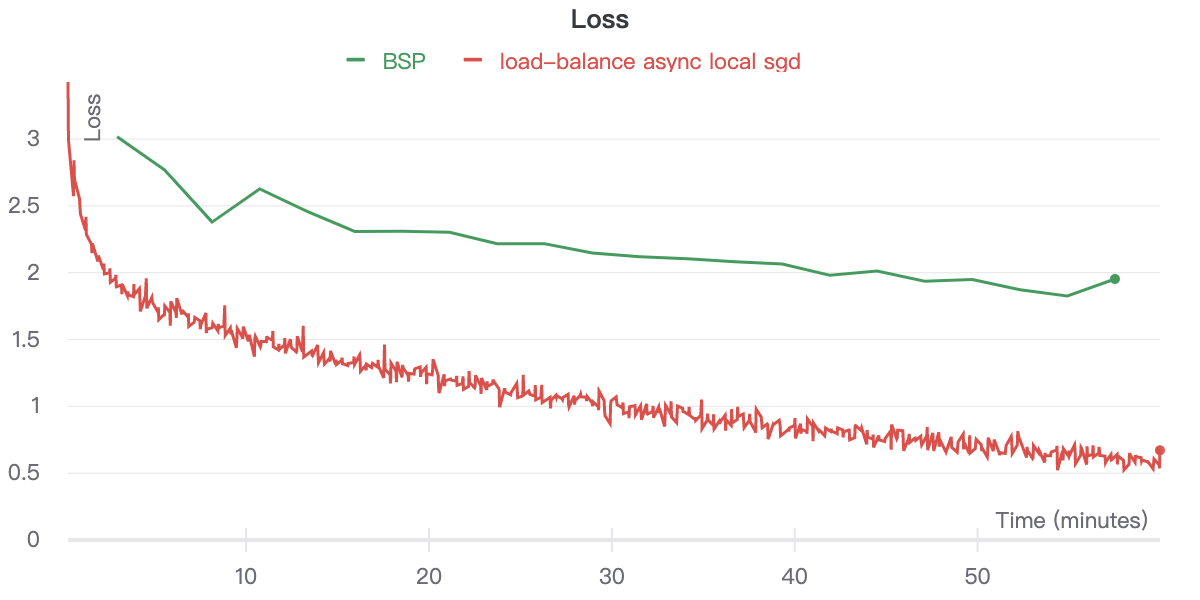}
    \caption{}
    \label{local_with_balancing}
\end{figure}
\begin{figure}[ht]
    \centering
    \includegraphics[width=1\columnwidth]{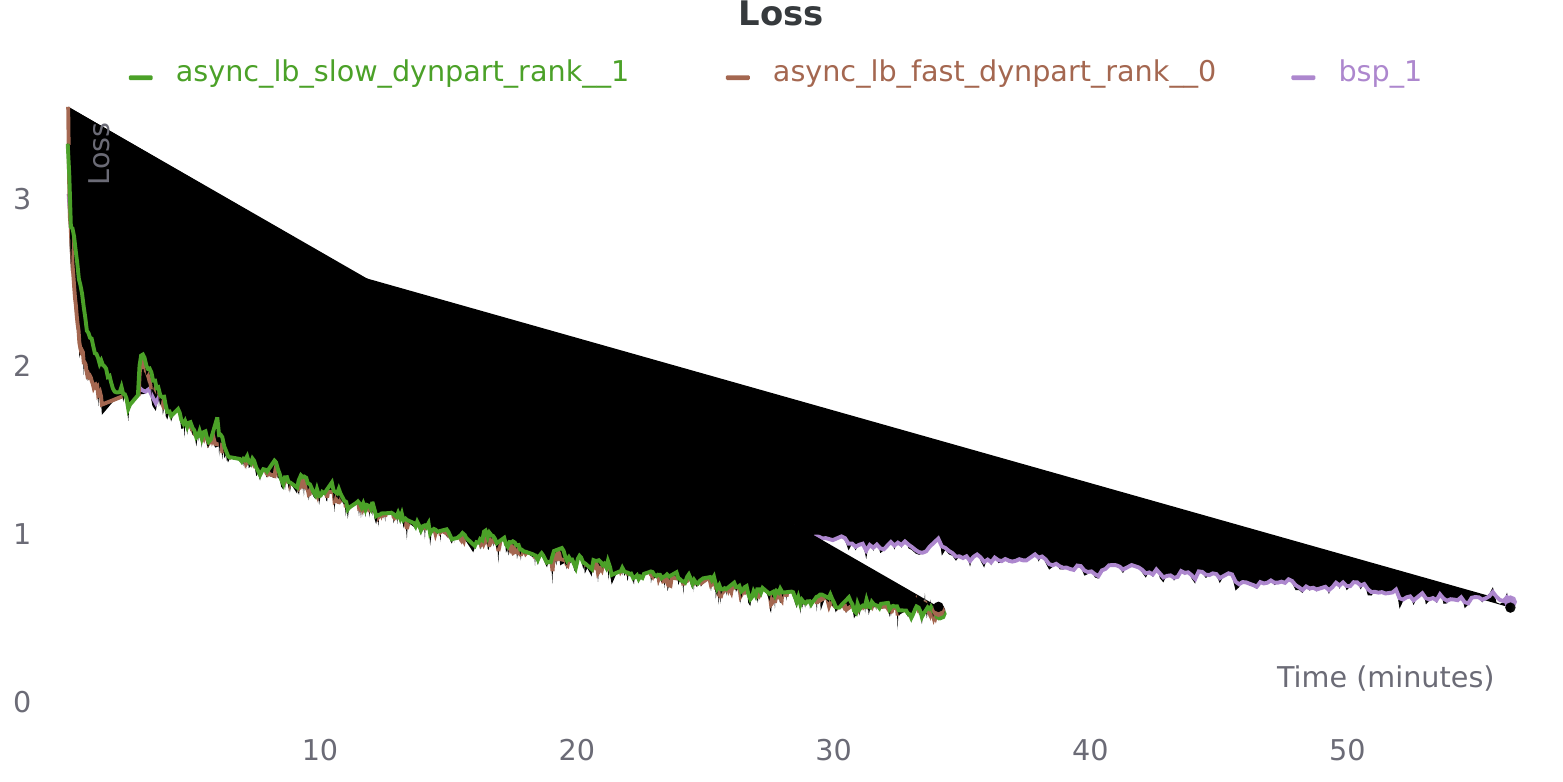}
    \caption{}
    \label{local_with_balancing}
\end{figure}

\begin{figure}[ht]
    \centering
    \includegraphics[width=1\columnwidth]{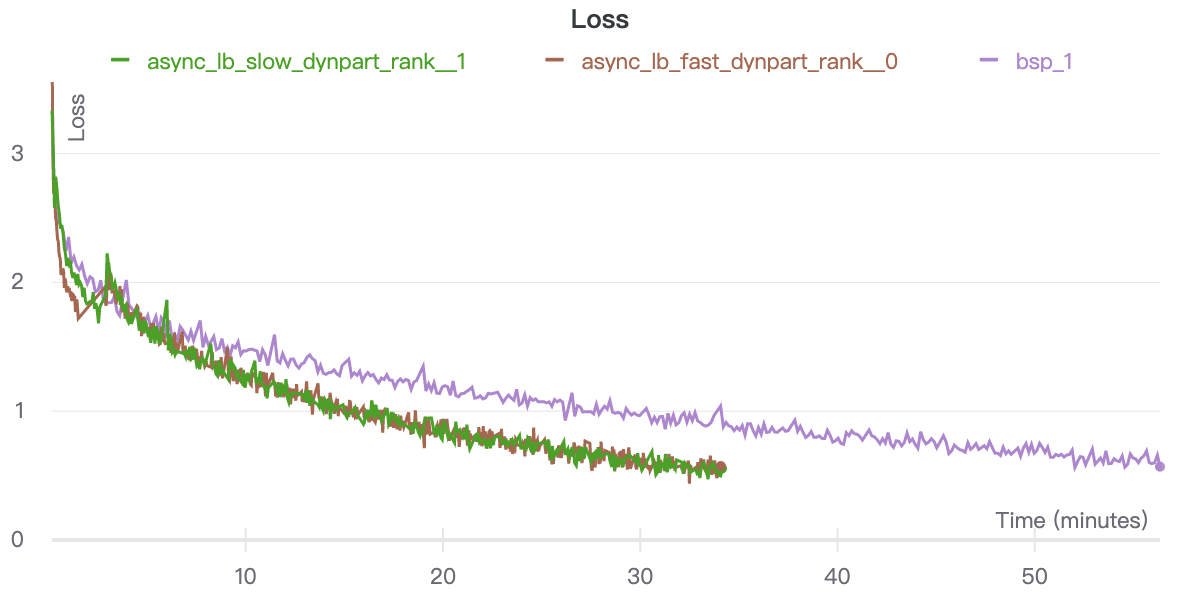}
    \caption{}
    \label{local_with_balancing}
\end{figure}

\subsection{Rate of convergence}
Figure 6 plots the training time in ResNet101. We set the training time to 1 hour, batchsize to 128. It can be clearly observed in the above figure that the curve of BSP is more smoother. This is due to the strong synchronization characteristics of BSP, which can ensure the correctness of the gradient from different workers and avoid shocks. Although the convergence process of BSP is very stable, its convergence speed is very slow. This is because in a heterogeneous environment, the performance of each worker is different, which causes different workers to process data of the same size in different times. 
Thus some workers with good performance are idle for most of the time. Our approach uses load balancing to improve the utilization of computing resources, so that more data can be iterated in the same time. Thereby accelerating convergence.
\subsection{Scalability}

\section{Related Works}
\label{rel}
\subsection{Asynchronous SGD}
\label{async}

For large scale machine learning optimization problems, parallel mini-batch SGD suffers from synchronization delay due to a few slow machines, slowing down entire computation. To mitigate synchronization delay, asynchronous SGD method are studied in \cite{recht2011hogwild,de2015taming,lian2015asynchronous}. These methods, though faster than synchronized methods, lead to convergence error issues due to stale gradients. \cite{agarwal2011distributed} shows that limited amount of delay can be tolerated while preserving linear speedup for convex optimization problems. Furthermore, \cite{zhou2018distributed} indicates that even polynomially growing delays can be tolerated by utilizing a quasilinear step-size sequence, but without achieving linear speedup.

\subsection{Large batch SGD}
\label{large}

Recent schemes for scaling training to a large number of workers rely on standard mini-batch SGD with very large overall batch sizes \cite{you2018imagenet,goyal2017accurate} , i.e. increasing the global batch size linearly with the number of workers K. \cite{yu2019computation} has shown that remarkably, with exponentially growing mini-batch size it is possible to achieve linear speed up (i.e., error of \(\mathcal{O}(1/KT))\) with only \(\log{T}\) iterations of the algorithm, and thereby, when implemented in a distributed setting, this corresponds to \(\log{T}\) rounds of communication. The result of \cite{yu2019computation} implies that SGD with exponentially increasing batch sizes has a similar convergence behavior as the full-fledged (non-stochastic) gradient descent.

While the algorithm of \cite{yu2019computation} provides a way of reducing communication in distributed setting, for a large number of iterations, their algorithm will require large minibatches, and washes away the computational benefits of the stochastic gradient descent algorithm over its deterministic counter part. Furthermore, it has been found that increasing the mini-batch size often leads to increasing generalization errors, which limits their distributivity \cite{li2014efficient}.

Our work is complementary to the approach of \cite{yu2019computation}, as we focus on approaches that use local updates with a fixed minibatch size, which in our experiments, is a hyperparameter that is tuned to the data set.

\subsection{Local SGD}
\label{local}

Motivated to better balance the available system resources (computation vs. communication), local SGD (a.k.a. local-update SGD, parallel SGD, or federated averaging) has recently attracted increased research interest \cite{zinkevich2010parallelized,mcdonald2010distributed,zhang2014improving,mcmahan2017communication}. In local SGD, each worker evolves a local model by performing H sequential SGD updates with mini-batch size B, before communication (synchronization by averaging) among the workers.

A main research question is whether local-update SGD provides a linear speedup with respect to the number of workers \(K\), similar to mini-batch SGD. Recent work partially confirms this, under the assumption that \(H\) is not too large compared to the total iterations \(T\). \cite{stich2018local} show convergence at \(\mathcal{O}((KT)^{-1})\) on strongly convex and smooth objective functions when \(H = \mathcal{O}(T^{1/2})\). For smooth non-convex objective functions, \cite{yu2019parallel} give an improved result \(\mathcal{O}((KT)^{-1/2})\) when \(H = \mathcal{O}(T^{1/4})\). \cite{zhang2016parallel} empirically study the effect of the averaging frequency on the quality of the solution for some problem cases and observe that more frequent averaging at the beginning of the optimization can help. Similarly, \cite{bijral2016data} argue to average more frequently at the beginning.

Although existing works provides convergence guarantees on local-update SGD, there is still no effort focus on optimally tuning local-update SGD to heterogeneous settings (slower workers do less computation between synchronization, and faster workers do more) using load-balancing techniques.

\section{Conclusion}
\label{con}
This is Conclusion.

\bibliography{main}
\end{document}